\documentclass[a4paper,11pt]{article}
\pdfoutput=1
\usepackage{jheppub}
\usepackage{graphicx}
\usepackage{amsmath}
\usepackage{amsfonts}
\usepackage{amssymb}
\usepackage{slashed}
\usepackage[colorlinks=true,linkcolor=blue]{hyperref}
\usepackage{bm}
\usepackage{empheq}
\usepackage[super]{nth}

\def\half{\frac{1}{2}}

\def\LL{\mathcal{L}}
\def\MM{\mathcal{M}}
\def\C{\mathcal{C}}
\def\be{\begin{equation}}
\def\ee{\end{equation}}
\def\bea{\begin{eqnarray}}
\def\eea{\end{eqnarray}}
\def\Eq#1{Eq.~\eqref{#1}}
\def\Re{\,\mathrm{Re}\,}

\def\Pii{\Pi_{\mbox{\scriptsize{1-order}}}}
\def\GTT{G^{\pi\pi}_{\mathrm{s}}}
\def\GTTR{G^{\pi\pi}_{\mathrm{R}}}
\def\p{{\bm{p}}}
\def\k{{\bm{k}}}

\title{Stress-stress correlator in $\phi^4$ theory:  Poles or a Cut?}
\author{Guy D.\ Moore}

\affiliation{Institut f\"ur Kernphysik, Technische Universit\"at Darmstadt\\
Schlossgartenstra{\ss}e 2, D-64289 Darmstadt, Germany}
\emailAdd{guy.moore@physik.tu-darmstadt.de}

\abstract{
We explore the analytical properties of the traceless stress tensor
2-point function at zero momentum and small frequency (relevant for
shear viscosity and hydrodynamic response) in hot, weakly
coupled $\lambda \phi^4$ theory.  We show that, rather than one or a
small number of poles, the correlator has a cut along the negative
imaginary frequency axis.  We briefly discuss this result's relevance
for constructing 2'nd order hydrodynamic models of hot relativistic
field theories.
}

\begin{document}
\maketitle
\section{Introduction}
\label{sec:intro}

Relativistic hydrodynamics is the universal effective theory for
describing relativistic fluids which vary on sufficiently long time
and length scales that they remain locally near equilibrium (even if
the temperature, velocity, and any chemical potentials vary by a large
amount on long scales, for a comprehensive recent review see
\cite{Schafer:2009dj}).  For real-world applications it is important
to include the effects of small departures from local equilibrium,
arising from the fluid's nonuniformity, see \cite{Teaney:2009qa} for a
review.  Often this disequilibrium is controlled primarily by the shear
viscosity $\eta$.  But a \textsl{relativistic} hydrodynamic theory
containing only shear viscosity suffers from instability and
acausality problems \cite{Muller:1967zza,Hiscock:1983zz,Hiscock:1985zz},
and must be supplemented with additional, higher-order
corrections \cite{Israel:1976tn,Israel:1979wp}.  Qualitatively these
take the form of an exponential relaxation of the stress tensor
towards a viscous-fluid form.  They lead to a more complex
hydrodynamic theory described by a larger number of coefficients.

There is an ongoing discussion on how to correctly interpret
higher-order hydrodynamics.  In the absence of conserved quantities
besides energy and momentum, hydrodynamics is a theory to determine
the evolution of the stress tensor $T^{\mu\nu}$ from its equation of
state, $P=P(\varepsilon)$ with $P$ the pressure and $\varepsilon$ the
energy density (each in the local rest frame).  Shear viscosity enters
the first-order theory as a correction to the functional form of the
stress tensor,%
\footnote{We are using $[{-}{+}{+}{+}]$ metric conventions.}
\begin{eqnarray}
  \label{order1}
  T^{\mu\nu} &=& T^{\mu\nu}_{\mathrm{eq}} + \Pi^{\mu\nu} \,, \quad
  T^{\mu\nu}_{\mathrm{eq}} =
  (\varepsilon{+}P) u^\mu u^\nu + P g^{\mu\nu} ,
  \\
  \label{Pii}
  \Pii^{ij} &=& -\eta \left( \partial^i u^j + \partial^j u^i
  - \frac{2}{3} \delta^{ij} \partial_k u^k \right)
  -\zeta \delta^{ij} \partial_k u^k \,,
\end{eqnarray}
where $T^{\mu\nu}$ is the stress tensor, $u^\mu$ is the flow
4-velocity in the Landau-Lifshitz frame ($u_\mu \Pi^{\mu\nu}=0$), and
the expression for the first-order nonequilibrium stress
$\Pii^{\mu\nu}$ has been written for the local rest frame.  The
first-order theory consists of using $\Pii^{ij}$ from \Eq{Pii}  as
$\Pi^{ij}$ in \Eq{order1}.
The expression shows that the 5 $\ell{=}2$ modes behave quite
differently than the one $\ell{=}0$ ($\delta^{ij}$) mode, so we
implicitly project out the $\delta^{ij}$ component of $T^{ij}$ to
concentrate on the $\ell=2$ part in the following; that is, from now
on, when we write $T^{ij}$ we mean $T^{ij} - \delta^{ij} T^{kk}/3$.

\Eq{order1} should be understood as an infrared effective description.
But for numerical applications we need to reformulate \Eq{order1} in a way
which leads to stable evolution on all scales; the reformulation must
reduce to $\Pi^{ij} = \Pii^{ij}$ for slowly-varying systems, and if
the reformulation leads to more accurate behavior that is an added
benefit.  Israel and Stewart proposed a now-standard second-order
reformulation \cite{Israel:1976tn,Israel:1979wp}, in which $\Pi^{ij}$
is related to $\Pii^{ij}$ through a relaxation process,
\begin{equation}
  \label{IS1}
  \tau_\pi \partial_t \Pi^{ij} = \Pii^{ij}
  - \Pi^{ij} \,.
\end{equation}
This introduces a new coefficient $\tau_\pi$.  In one interpretation,
this coefficient should be chosen in order to optimize the accuracy of
$\Pi^{ij}$ in a slowly varying system.  That is, the transport
coefficient $\tau_\pi$, and others which appear at this order in
derivatives, should be chosen so as to optimally describe the behavior
of hydrodynamic systems which vary slowly in space and time.
This approach is implicit for instance by Baier \textsl{et al}
\cite{Baier:2007ix} (see also \cite{Bhattacharyya:2008jc}),
who use it to derive a Kubo relation for this coefficient, which is
then evaluated in a strongly coupled holographic theory
\cite{Baier:2007ix}.

Alternatively, we can interpret \Eq{IS1} as an attempt to really
describe the microscopic physics by which the off-equilibrium stress
tensor $\Pi^{ij}$ approaches its near-equilibrium form
$\Pii^{ij}$.  In equilibrium in the absence of flow such that
$\Pii^{ij}=0$, the initial $\Pi^{ij}(t=0)$ can be interpreted as that
due to random thermal fluctuations, and the equation then gives a
specific prediction for the stress tensor autocorrelator.  \Eq{IS1}
amounts to an \textsl{Ansatz} that the two-point correlator is
controlled by a retarded function with a single pole%
\footnote{%
  \Eq{TTcorr} doesn't have a single pole; it has two poles, at
  $\omega = \pm i\tau_\pi^{-1}$. That is because it is the symmetrized
  correlator, not the retarded one; we will explain the relation
  shortly.}
with imaginary part $-i/\tau_\pi$,
\begin{eqnarray}
  \label{TTcorr}
 \int d^3 x \langle \Pi^{ij}(x,t) \Pi^{ij}(0,0) \rangle
  & = & e^{-|t|/\tau_\pi}  \int d^3 x \langle \Pi^{ij}(x,0) \Pi^{ij}(0,0) \rangle
  \; \Rightarrow \; \\
  \GTT(\omega) & \equiv &
  \int e^{i\omega t} \langle \Pi^{ij}(x,t) \Pi^{ij}(0,0) \rangle  d^3x \: dt
  \nonumber \\
  & = & 10T \frac{4P}{5} \int e^{i\omega t} e^{-|t|/\tau} dt
  = 8PT
  \frac{2 \tau_\pi^{-1}}{\omega^2 + \tau_\pi^{-2}} \,. \nonumber
\end{eqnarray}
Here $10$ is the number of $\ell=2$ components appearing in the sum in
$T^{ij} T^{ij}$ and $4PT/5$ is the equal-time, mean-squared
fluctuation in each component in our kinetic description.%
\footnote{%
  The number of terms is 10 because there are $2\ell+1=5$ independent
  terms which are each double-counted in the sum, for instance,
  $T^{xy}=T^{yx}$ is one of the 5 terms but both $T^{xy} T^{xy}$ and
  $T^{yx} T^{yx}$ appear in the sum.  The equal-time correlator
  $\int d^3 x \langle \Pi^{xy}(x,0) \Pi^{xy}(0,0) \rangle$
  equals the $\omega$-integral of $\GTT(\omega)/10$ and should equal
  $PT$.  At weak coupling, $4/5$ of this contribution arises from very
  small frequencies which are accounted for in this kinetic theory
  calculation, and $1/5$ arises from cut-type structures at
  frequencies $\omega \sim T$.  Hence the factor $4/5$ in our
  expression.}
This is exactly the behavior of the correlator in the so-called
relaxation-time approximation.

Denicol and collaborators have argued \cite{Denicol:2011fa}
that we can use a slightly modified form of \Eq{IS1} to improve the
behavior of a slowly varying system \textsl{and} to simulate with
maximal fidelity the microscopic behavior, by replacing \Eq{IS1} with
\begin{equation}
  \label{IS2}
  \tau_{\pi,\mathrm{micro}} \, \partial_t \Pi^{ij} = -\Pi^{ij}
  + \left( \Pii^{ij}
  + (\tau_{\pi,\mathrm{micro}}-\tau_{\pi,\mathrm{macro}}) \,
  \partial_t  \Pii^{ij} \right) \,,
\end{equation}
where $\tau_{\pi,\mathrm{macro}}$ is the
definition in terms of slowly varying systems and Kubo relations, and
$\tau_{\pi,\mathrm{micro}}$ is the imaginary part of the pole in
$\GTT(\omega)$ closest to the real axis.
This definition assumes (as noted in \cite{Denicol:2011fa}) that the
stress tensor Green function has such a pole, rather than a cut
structure.  This appears to be a reasonable assumption.  For instance
it is consistent with the behavior obtained in strongly-coupled
analogue theories with holographic duals, such as $\mathcal{N}{=}4$
SYM theory at strong coupling with many colors.  In this case we know
that the dual theory contains a black hole, and the relevant
correlator shows exponential decay, characterized by several complex
exponents determined by the quasinormal modes of this dual black hole
\cite{Kovtun:2005ev}.  Note however that the behavior of such analogue
theories at finite coupling, where known, is generally more complex
\cite{Grozdanov:2016vgg}; the poles with $\Re \omega \neq 0$ move
towards the real axis, and new poles appear on the imaginary axis and
move towards the origin.%
\footnote{%
  Note that the $\langle T^{ij} T^{ij}\rangle$ correlator is called
  the scalar channel in the holography community; they reserve ``shear
  channel'' to describe $\langle T^{0i} T^{0i} \rangle(k)$ with $i$ unsummed
  and orthogonal to spatial $\k$.}
The relaxation-time approximation, which is often considered for
simplicity and has sometimes been advocated on theoretical grounds
\cite{Koide:2009sy,Koide:2008nw}, corresponds to the presence of
exactly one pole in the lower half-plane.

But it is by no means obvious that $\GTT$ really is controlled by
one or a small number of poles.  It would be useful to know the
behavior of more theories to see whether they contain poles, like SYM
theory, or cuts.  In this paper we address this for weakly coupled
scalar $\lambda \phi^4$ theory.  We do so because the interactions are
simple enough to allow an extremely precise study within kinetic
theory, which is the relevant effective description for small
self-coupling $\lambda$.  We expect this theory to be representative
of the behavior of weakly coupled QCD.  If we see behavior
substantially different than in SYM theory, it raises interesting
questions about whether the analytic behavior of the 2-point function
really is controlled by poles.

The question we address bears some
similarity to the topic of two recent studies, one by
Romatschke \cite{Romatschke:2015gic} and one by Kurkela and Wiedemann
\cite{Kurkela:2017xis}.  However the emphasis is somewhat different.
These references wanted to study the wave-number $k$ dependence of
stress-stress correlation functions, including those which contain
hydrodynamic poles, whereas we will look only at the $k=0$ limit and
we concentrate on a correlator which determines a hydrodynamic
coefficient but does not itself possess hydrodynamic poles.  Because
of their broader subject, the other studies made simplifying
assumptions about scattering (relaxation-time approximations), whereas
the point of our study is precisely to avoid this and to consider the
full structure of scattering in a specific theory.  We will return to
this issue when we present our main results.  Both sets of studies
are relevant in comparing between weak-coupling behavior and the
structure of poles observed at strong coupling \cite{Kovtun:2005ev},
but our main motivation lies in the interpretation of second-order
hydrodynamics as explained above.  We therefore consider
the studies by Romatschke and by Kurkela and Wiedemann to be
complementary to ours.

In the next section we develop the tools to study the spectral
properties within kinetic theory approximately.  The approximate
procedure we have available always results in a $\GTT(\omega)$ which
is a rational function; therefore it is similar to the problem of
fitting analytic functions with a Pad\'e approximant.
So we will also give a quick sketch of what functions
with poles, and with cuts, look like when we try to fit them using
rational functions.  Finally we will present our results and argue
that they indicate the spectral function to have a cut, rather than a
series of well-separated (quasinormal mode) poles.  However, we will
show that the cut carries most of its spectral weight over a
relatively narrow range of frequencies.

\section{Calculation ingredients}
\label{sec:ingredients}

Here we review the tools which we will use to study the analytic
property of the $\GTT$ correlator in weakly-coupled theories via
kinetic theory.  We begin by reminding the reader of the relation
between retarded and symmetrized correlation functions.  Then we show
how the symmetrized correlation function is evaluated in kinetic
theory, and how kinetic theory is solved as a variational problem.

\subsection{Retarded and symmetrized correlators}
\label{sub:corr}

The Kubo relation for the shear viscosity is most naturally derived in
terms of the retarded correlation function of the stress tensor,
\begin{eqnarray}
  \label{eta_GR}
  \eta & = & \frac{1}{10} \lim_{\omega \to 0}
  \frac{1}{\omega} \, \mathrm{Im} \int d^4 x
  e^{i\omega t} \GTTR(x,t) \,, \\
  \label{GR}
  \GTTR(x,t) & \equiv &
  i \left\langle \Big[ \Pi^{ij}(x,t) \,,\,
  \Pi^{ij}(0,0) \Big] \right\rangle \Theta(t) \,.
\end{eqnarray}
Because of the $\Theta(t)$ factor, the retarded correlator
$\GTTR(\omega)$ is nonsingular for frequencies with nonnegative
imaginary part and has its singularities strictly in the lower
frequency half-plane.  It has a KMS relation with the symmetrized
correlator
\begin{equation}
  \label{KMS}
  \GTT(\omega) = \frac{e^{\omega/T}+1}{2(e^{\omega/T}-1)}
  \,\mathrm{Im}\: \GTTR(\omega)
  \simeq \frac{T}{\omega} \,\mathrm{Im}\: \GTTR(\omega) \,,
\end{equation}
where in the second step we have made a small-$\omega$ approximation,
which is appropriate at weak coupling where the relevant frequencies
will be suppressed by powers of the coupling $\omega \sim \lambda^2 T$.
Using this relation, a simple pole in the retarded function
corresponds to a pair of poles in the symmetrized correlation
function:
\begin{equation}
  \label{onepole}
  \GTTR(\omega) = \ldots + \frac{C}{\Gamma - i \omega}
  \quad \Rightarrow \quad
  \GTT(\omega) = \ldots + \frac{CT}{(\Gamma+i\omega)(\Gamma-i\omega)}
  \,.
\end{equation}
Though it is not our emphasis in this work, we remark that the
KMS relation lets us express the viscosity in terms of $\GTT$:
\begin{equation}
  \label{eta_GS}
  \eta = \frac{1}{10 T} \lim_{\omega \to 0} \int d^4 x \;
  e^{i\omega  t} \GTT(x,t) = \frac{1}{10 T} \int d^4 x \; \GTT(x,t) \,.
\end{equation}

Within kinetic theory, the stress tensor is expressed in terms of the
statistical function $f(p,x,t)$ as
\begin{equation}
  \label{Tkinthy}
  T^{ij}(x,t) =  \int \frac{d^3 p}{(2\pi)^3}
  \frac{p^i p^j}{E_p} f(p,x,t) \,, \qquad
  E_p \equiv \sqrt{p^2+m^2} \,.
\end{equation}
(In a multicomponent or multiparticle theory we should add an index
for particle type, which is summed over.  We will leave this out as we
will eventually specialize to a one-component scalar.)
For the $\ell=2$ components of the stress tensor, the angular average
above gives zero if we evaluate it using the equilibrium values of
$f$, $f_0=(\exp(-E_p/T)\mp 1)^{-1}$ for a system at rest at
temperature $T$.  However the statistical functions possess
fluctuations; the well-known Bose or Fermi number fluctuations are in
this context
\begin{equation}
  \label{ffcorr}
  \langle f(p,x,0) f(q,x',0) \rangle
  = f_0(p) f_0(q) + (2\pi)^3 \delta^3(p-q) \delta^3(x-x')
   f_0(p) [1{\pm} f_0(p)]
\end{equation}
with $\pm$ a $+$ for bosons and a $-$ for fermions, and with $f_0(p)$
the equilibrium mean occupancy.  Therefore
\begin{equation}
  \label{TTcorr2}
  \int d^3 x  \langle \Pi^{ij}(x) \Pi^{ij}(0) \rangle
  = \int \frac{d^3 p}{(2\pi)^3}
  \frac{p^i p^j - \frac{\delta^{ij} p^2}{3}}{E}
  \frac{p^i p^j - \frac{\delta^{ij} p^2}{3}}{E}
  f_0(p)[1{\pm}f_0(p)]
  = 10 \frac{4PT}{5} 
\end{equation}
with $P,T$ the pressure and temperature as before.  To determine the
time structure of this correlation function, we need to establish the
time dependence of the statistical function $f(x,p,t)$.

\subsection{Kinetic theory setup}
\label{sub:kin}

We consider scalar field theory with a single real field $\phi$ with
Lagrangian
\begin{equation}
  \label{Lagrangian}
  -\LL[\phi,\partial_\mu \phi] = \half \partial_\mu \phi
  \partial^\mu \phi + \frac{m^2}{2} \phi^2
  + \frac{\lambda}{24} \phi^4 \,,
\end{equation}
and in this work we consider $m^2 \ll T^2$ so it can be neglected and
we work perturbatively in $\lambda$.  The scattering matrix element is
precisely $\MM=\lambda$, which will lead to simple momentum dependence
in what follows.

At weak coupling and for the dominant $p \sim \pi T$ modes which
control thermodynamics and viscosity, we may make the quasiparticle
approximation and apply kinetic theory.  We will not review kinetic
theory in detail, referring the reader for its derivation and
application towards this problem to the literature
\cite{kadanoff1962quantum,Jeon:1994if,Jeon:1995zm,Arnold:1998cy,%
  Arnold:2000dr,Arnold:2003zc}.
The statistical function obeys a Boltzmann equation, which for our
space-uniform system is of the form
\begin{eqnarray}
  \label{Boltzmann}
  \hspace{-2em}
  \frac{\partial}{\partial t} f(\p,x,t) & = & - \C[f]
  \nonumber \\
  \C & = & \frac{1}{2p} \half
  \int \frac{d^3 k d^3 p' d^3 k'}   {(2\pi)^9 2k 2p' 2k'}
  (2\pi)^4 \delta^4(P{+}K{-}P'{-}K') |\MM|^2
  \nonumber \\
  &&{} \times \Big( f(\p) f(\k) [1{\pm} f(\p')][1{\pm} f(\k')
    \, - \, f(\p') f(\k') [1{\pm} f(\p)][1{\pm} f(\k) \Big) \,,
\end{eqnarray}
where the first line is the Boltzmann equation with $\C$ the collision
operator; the second line is the form of the collision operator for
$2 \leftrightarrow 2$ scattering; and the final line is the
combination of statistical functions showing removal and addition of a
particle of momentum $p$ (in which we have suppressed writing the
dependence on $(x,t)$).  Capital letters are 4-vectors while lower
case letters represent momenta or their magnitudes.  The factor $1/2$
in the first line of $\C$ is a final-state symmetry factor.  We are
interested in the case where a $t=0$ initial fluctuation is of form
(the factor $\sqrt{3/2}$ is to follow the conventions of
\cite{Arnold:2000dr})
\begin{equation}
  \label{chi}
  \delta f(\p,t=0) = X_{ij} \sqrt\frac{3}{2}
  \frac{p^i p^j - \frac{\delta^{ij} p^2}{3}}{E}
  f_0(p) [1{\pm} f_0(p)]
\end{equation}
so that a specific component $\Pi^{ij} \propto X^{ij}$ is nonzero; we
then want to see how that fluctuation relaxes with time, so we can
convert the equal-time correlator in \Eq{TTcorr2} into an unequal time
correlator and then into a frequency-domain correlator.  Alternatively
in the case that the fluid is under shear flow, $X_{ij}$ represents
the shear stress applied on the fluid \cite{Arnold:2000dr}.  The
angular dependence is captured by writing $\delta f(\p,t)$
with the \textsl{Ansatz}
\begin{equation}
  \delta f(\p,t) = X_{ij} \chi_{ij}(\p,t) f_0(p) [1{\pm} f_0(p)]
  \,, \qquad
  \chi_{ij}(\p,t) = \sqrt{3/2} \Big(\hat p_i \hat p_j - \delta_{ij}/3 \Big)
  \chi(p,t) \,.
\label{df-is}
\end{equation}
The form of the \textsl{Ansatz} is ensured by the rotational symmetry
of the theory and the fact that we work only to linear order in
perturbations.  To find the time evolution of $\chi(p,t)$ we insert
\Eq{df-is} in \Eq{Boltzmann}, finding after a little work
\cite{Jeon:1995zm}
\begin{eqnarray}
  - \frac{\partial \chi(p,t)}{\partial t} & = &
  \C[\chi(p,t)] \nonumber \\
  & = & \frac{\lambda^2}{2} \int \frac{d^3 k d^3 p' d^3 k'}
    {(2\pi)^9 2p 2p' 2k 2k'} (2\pi)^4 \delta^4(P{+}K{-}P'{-}K')
    f_0(p) f_0(k) [1{+}f_0(p')][1{+}f_0(k')]
    \nonumber \\ && {} \times
    \Big( \chi(p,t) + P_2(c_{pk}) \chi(k,t)
    - P_2(c_{pp'}) \chi(p',t) - P_2(c_{pk'}) \chi(k',t) \Big) \,.
\label{Boltzlinear}
\end{eqnarray}
Here $P_2(c_{pk})$ is the second Legendre polynomial with
$c_{pk} = \cos \hat{p}\cdot \hat{k} = \vec p \cdot \vec k / pk$.

Formally, the space of possible $\chi(p)$ form an infinite dimensional
vector space $\LL^2$ (Lebesgue-square-integrable functions) with
measure
\bea
  \label{measure}
  \langle \chi_1(p) \, | \, \chi_2(p) \rangle & \equiv &
  T^{-3} \int \frac{d^3 p}{(2\pi)^3} \chi_{ij1}(\p) \chi_{ij2}(\p)
  f_0(p) [1{\pm} f_0(p)]
  \\ \nonumber & = &
  \frac{4\pi}{(2\pi T)^3} \int f_0(p) [1{\pm} f_0(p)]
  \chi_1(p) \chi_2(p) \: p^2 \, dp
\eea
and $\C(\chi)$ acts as a positive symmetric operator on this space
\cite{Arnold:2000dr}.  As such, it can be expressed in terms of its
spectrum, which may contain both discrete and continuous components.
Specifically, we can write the action of the collision operator in
terms of eigenvalues and eigenvectors as
\begin{equation}
  \label{C_acts}
  \C \, | \, \chi \rangle =
  \left( \sum_i \lambda_i \,|\,\xi_i \rangle \langle \xi_i \,|
  + \int d\lambda_j \: \lambda_j \,|\, \xi_j \rangle \langle \xi_j\,|
  \right) \, | \, \chi \rangle
\end{equation}
with $\lambda_i$ the discrete eigenvalues with eigenvectors
$|\,\xi_i \rangle$ and with the $\int d\lambda_j$ integration running
over any continuous spectrum the operator may possess.  The
eigenvectors are orthonormal,
$\langle \xi_{i_1} \,|\, \xi_{i_2} \rangle = \delta_{i_1 i_2}$ and
$\langle \xi_{j_1} \,|\, \xi_{j_2} \rangle = \delta(j_1-j_2)$
(Kroneker and Dirac delta functions).

In terms of this (in principle solvable) spectral decomposition, the
departure at an arbitrary time is solved as
\begin{equation}
  \label{Boltz_sol}
  \chi(p,t) = \sum_i \, \langle \xi_i\,|\, \chi(p,0) \rangle
  \: \xi_i(p) e^{-\lambda_i t}
  + \int d\lambda_j \: \langle \xi_j \,|\, \chi(p,0) \rangle
  \: \xi_j(p) e^{-\lambda_j t} \,.
\end{equation}
That is, the initial departure from equilibrium is decomposed in terms
of the eigenvectors of the collision operator, which each decay
exponentially at a rate controlled by the respective eigenvalue.
The Boltzmann equation should only be used at positive times because
it is structured in terms of a dissipative response to an initial
condition, but by time symmetry the negative-$t$ correlators are the
same as at positive $t$, and so the stress-stress correlator and the
shear viscosity evaluate to
\begin{eqnarray}
  \label{G_evec}
  \GTT(\omega) & = & \sum_i \frac{2\lambda_i T^3}{\lambda_i^2 + \omega^2}
  \left| \langle \chi(p,0) \,|\, \xi_i \rangle \right|^2
+ \int d\lambda_j \frac{2\lambda_j T^3}{\lambda_j^2 + \omega^2}
\left| \langle \chi(p,0) \,|\, \xi_j \rangle \right|^2 \,,
\\
\eta & = & \frac{\GTT(0)}{10T} =
\sum_i \frac{T^2}{5\lambda_i}
  \left| \langle \chi(p,0) \,|\, \xi_i \rangle \right|^2
+ \int d\lambda_j \frac{T^2}{5 \lambda_j}
\left| \langle \chi(p,0) \,|\, \xi_j \rangle \right|^2 \,.
\label{eta_evec}
\end{eqnarray}
This is then evaluated by treating the squared fluctuation
$|\, \chi(p,0) \rangle \langle \chi(p,0) \,|$ using \Eq{ffcorr}.
The eigenvalues of the discrete/continuous spectrum of the linearized
collision operator $\C$ correspond to locations of poles/cuts in the
stress tensor Green function, with residue/discontinuity determined by
the overlap of the associated eigenvector with the initial departure
from equilibrium.

\subsection{Variational solution}
\label{sub:var}

The collision operator is determined through its matrix elements.
Using the definitions of the previous section, we have
\bea
\label{Cij}
\langle \chi_1 |\, \C \, | \chi_2 \rangle  & \!\!=\!\! &
\int \frac{d^3 p d^3 k d^3 p' d^3 k'}{(2\pi)^{12} 2p 2k 2p' 2k'}
(2\pi)^4 \delta^4(P{+}K{-}P'{-}K') |\MM_{\p\k\p'\k'}|^2
\nonumber \\ && {} \times 
f_0(p) f_0(k)[1{\pm} f_0(p')] [1{\pm} f_0(k')]
\times \Big( \chi \mbox{-factor} \Big) \,,
\\
\label{chifactor}
\Big( \chi \mbox{-factor} \Big) \! & \!\!=\!\! &
\frac{1}{2} \chi_{ij1}(\p)
\Big[ \chi_{ij2}(\p) {+} \chi_{ij2}(\k)
    {-} \chi_{ij2}(\p') {-} \chi_{ij2}(\k') \Big]
    \\ & = & \nonumber
    \frac{1}{8}
    \Big[ \chi_{ij1}(\p) {+} \chi_{ij1}(\k)
    {-} \chi_{ij1}(\p') {-} \chi_{ij1}(\k') \Big]
\Big[ \chi_{ij2}(\p) {+} \chi_{ij2}(\k)
    {-} \chi_{ij2}(\p') {-} \chi_{ij2}(\k') \Big].
\eea
Here we use the symmetry of the first two lines to symmetrize the
$\chi_1$ dependence of the first factor.  To evaluate this we use
repeatedly that
\begin{equation}
  \label{Legendre}
  \chi_{ij1}(\p) \chi_{ij2}(\k)
  = P_2(c_{pk}) \chi_1(\p) \chi_2(\k) \,.
\end{equation}
Because $|\MM_{\p\k\p'\k'}|^2=\lambda^2$, we also use the
``s-channel'' integration variables \cite{Arnold:2003zc}
\bea
\label{Schannel}
\langle \chi_1 \,|\, \C \, | \, \chi_2 \rangle & = &
\frac{\lambda^2}{(4\pi)^6 T^3} \int_0^\infty d\omega
\int_0^\omega dp \int_0^\omega dp'
\int_{\omega-2\mathrm{min}[p,p',k,k']}^\omega dq
\int_0^{2\pi} d\phi
\\ \nonumber && {} \times
f_0(p) f_0(k) [1{+}f_0(p')][1{+}f_0(k')]
\\ \nonumber && {} \times
\Big( \chi_1(p) \chi_2(p) P_2(c_{pp})
+ \chi_1(p)\chi_2(k) P_2(c_{pk}) - \chi_1(p)\chi_2(p') P_2(c_{pp'})
+ \ldots \Big),
\eea
where the final line contains 16 total terms corresponding to each
$\chi_1$ and $\chi_2$ argument ranging over $(p,k,p',k')$; the sign is
positive/negative for an even/odd
number of primes (final-state particles).  Here $k,k'$ are
\begin{equation}
\label{kk'}
  k = \omega - p \,,  \qquad  k' = \omega - p' \,,
\end{equation}
the Mandelstam variables are  
\begin{equation}
\label{stu}
  s = \omega^2 - q^2 \,, \quad u = -s-t \,, \quad
  t = \frac{s}{2q^2} \Big( (p{-}k)(p'{-}k') - q^2
  +\cos\phi \sqrt{(4pk{-}s)(4p'k'{-}s)} \Big) \,,
\end{equation}
and the cosines of angles are $c_{pp} = 1 = c_{p'p'} = \cdots$ and
\begin{align}
  c_{pk}  & = 1 - \frac{s}{2pk} \,, & c_{p'k'} & = 1-\frac{s}{2p'k'} \,, &
  c_{pp'} & = 1 + \frac{t}{2pp'} \,, \nonumber \\
  c_{kk'} & = 1 + \frac{t}{2kk'} \,, & c_{pk'} & = 1+\frac{u}{2pk'} \,, &
  c_{p'k} & = 1 + \frac{u}{2p'k} \,.
  \label{allangles}
\end{align}
The $\phi$ integral is trivial and the $q$ integral can be performed
as well because no statistical or $\chi$-function depends on it.
However we are only able to accomplish the remaining integrals
numerically even if the forms of $\chi_1$ and $\chi_2$ are known.
For this reason, to date we have not been able to explicitly solve the
eigenvalue/eigenvector decomposition of $\C$ without further
approximation.  Note that so far our only approximation has
been the small $\lambda$ expansion.

At this point we abandon an exact eigenvalue/eigenvector decomposition
of $\C[\chi]$ and attempt such a decomposition only within a
restricted eigenspace, by requiring $\chi(p)$ to lie within a linear
\textsl{Ansatz}.  This is equivalent to replacing the full Hilbert
space $\LL^2$ with the Hilbert subspace spanned by the chosen
\textsl{Ansatz} eigenfunctions.  For a well chosen and flexible basis
of functions, this will typically capture the most important functions
in the sense of those which dominate \Eq{G_evec}.  Relevant properties
should converge as we enlarge the considered basis, just as the
low-lying eigenvalues become more accurate as one uses the same
procedure to find the spectrum within the variational approach to
quantum mechanics.  But there are limitations when $\C$ has a
continuous spectrum, which we will address in the next section.

We choose the following variational form:
\begin{equation}
  \label{chiAnsatz}
  \chi(p) = \sum_{i=1}^{M} c_i \varphi_i(p) \,, \qquad
  \varphi_i(p) \equiv \frac{p^{i+1} T^{N-i-1}}{(p+p_0)^{N-1}} \,,  
\end{equation}
where $p_0$ is an energy scale and $M,N$ are integers which control
the size and form of the basis.  This is the same \textsl{Ansatz} used in
\cite{Jeon:1995zm,Arnold:2000dr} except that we don't restrict either
to $p_0=T$ or to $M=N$.  It spans rational functions with
$(p+p_0)^{N-1}$ denominator, which flexibly accommodates functional
forms with structure between about $p_0/N$ to $Np_0$.  The most IR
behavior is restricted to be $p^2$ because the collision integral
grows rapidly with small $p$ so that this functional dependence almost
always occurs.  We can determine how flexible the \textsl{Ansatz} is
by picking $M$, with larger values giving a more flexible functional
form.  We can vary \textsl{where} this flexibility occurs by varying
$p_0$, with larger values saving more of the functional freedom for
larger $p$ values.  And we can allow functions with stronger UV
behavior by allowing $M>N$.  The basis of $\varphi_i(p)$ functions is
not orthonormal but we make it so by applying the Gram-Schmidt process
using the inner product given in \Eq{measure}.  We then evaluate the
matrix form of $\C[\chi]$ by performing the integrals in 
\Eq{Schannel} by numerical quadratures.  We are able to get stable and
precise matrix elements and eigenvector decomposition for up to 20
basis elements.  As a cross-check, we use \Eq{eta_evec}
to evaluate $\eta$ for comparison with previous accurate
evaluations.  Checking for convergence with basis size and quadratures
integration refinement, we find
\begin{equation}
  \eta = 3033.5425 \frac{T^3}{\lambda^2} \,,
\end{equation}
which compares well with 3040 found in \cite{Jeon:1995zm} and
3033.54 found in \cite{Moore:2007ib}.

\section{Discrete approximations to continuous spectra}
\label{sec:discrete}

There is a problem with our procedure.  When we restrict to a
finite-element \textsl{Ansatz}, or equivalently we work within a
finite-dimensional subspace of the infinite-dimensional $\LL^2$ space
of $\chi(p)$ functions, we automatically modify the possible form of
the collision operator's spectrum.  This is familiar from quantum
mechanics.  In a finite-dimensional Hilbert space, Hermitian (for us,
real symmetric) operators automatically have a discrete spectrum.  But
in infinite-dimensional (but separable) Hilbert spaces, Hermitian
operators generically have both discrete and continuous spectra%
\footnote{%
  Technically in generality self-adjoint operators can have
  pure-point, singular, and absolutely continuous spectrum
  \cite{Reed:1972}.}.
The spectrum we find should go over to the infinite-dimensional
spectrum in the limit that we enlarge our basis without limit.  But
how does a discrete spectrum turn into a continuous one, and how does
one recognize whether that is what is happening?

\subsection{Example: Pad\'e approximation}

For this purpose we find it useful to look at another example.
Consider analytical functions, which can have poles, zeros, and cuts.
The process of approximating $\C$ with a finite-dimensional Hilbert
space is similar to rational approximations of analytical functions; a
rational function has only a finite number of potential poles and
zeros and cannot possess a cut.  But in the limit that we take a
rational approximation with more and more terms, it should converge to
the underlying analytical function, with its cut structure.  Therefore
we will make an aside to explore how this looks in a few examples, and
how one identifies what is converging to a pole and what to a cut in
the high-order limit.

Consider first the function $\ln(1+x)$.  It is terribly fit by its
Taylor series for $|x|>1$ because a Taylor series fits an analytical
function with a function possessing zeros but no poles.  The logarithm
function has a cut from $-1$ to $-\infty$, and a cut is better fit by
a nearly-equal number of zeros and poles.  Therefore it would be
better to approximate the function with an $(M,N)$ Pad\'e approximant
with $M=N$ or $M=N+1$.  We can make the Pad\'e approximant unique by
forcing the first $M+N$ terms in its Taylor series to match the Taylor
series of the logarithm function, resulting in
\begin{equation}
  \label{Padelog}
  P_{11}(x) = \frac{1+\frac{3}{2} x}{1+\frac{1}{2} x} \,, \quad
  P_{22}(x) = \frac{1+2x+\frac{2}{3} x^2}{1+x+\frac 16 x^2} \,, \quad
  P_{33}(x) = \frac{1+\frac 52 x + \frac 85 x^2 + \frac{7}{30} x^3}
  {1+\frac 32 x + \frac 35 x^2 + \frac{1}{20}x^3} \,, \ldots
\end{equation}
The zeros/poles of each Pad\'e approximant correspond to the
zeros of the polynomial in the numerator/denominator.

\begin{figure}[ht]
  \includegraphics[width=0.46\textwidth]{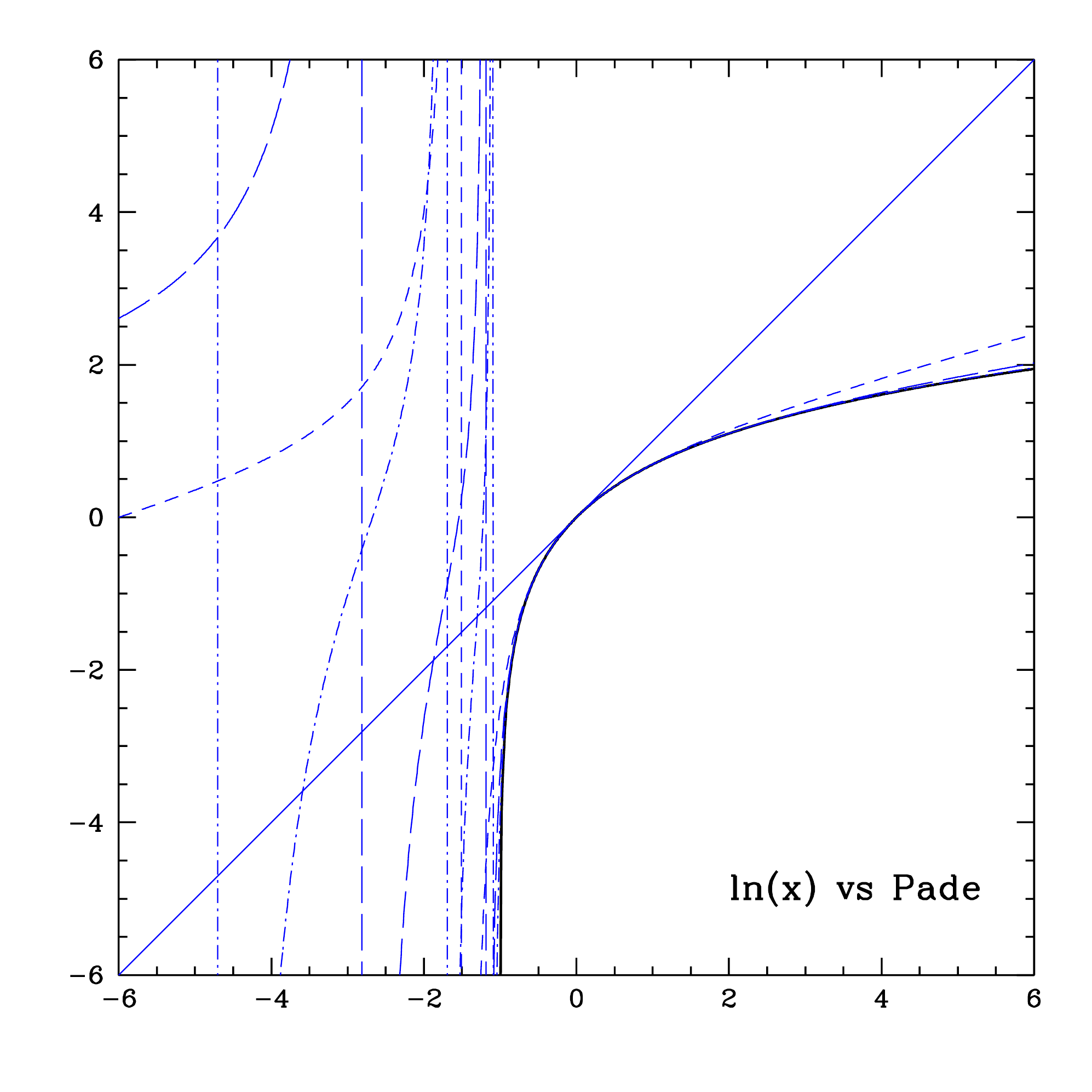}
  \hfill
    \includegraphics[width=0.46\textwidth]{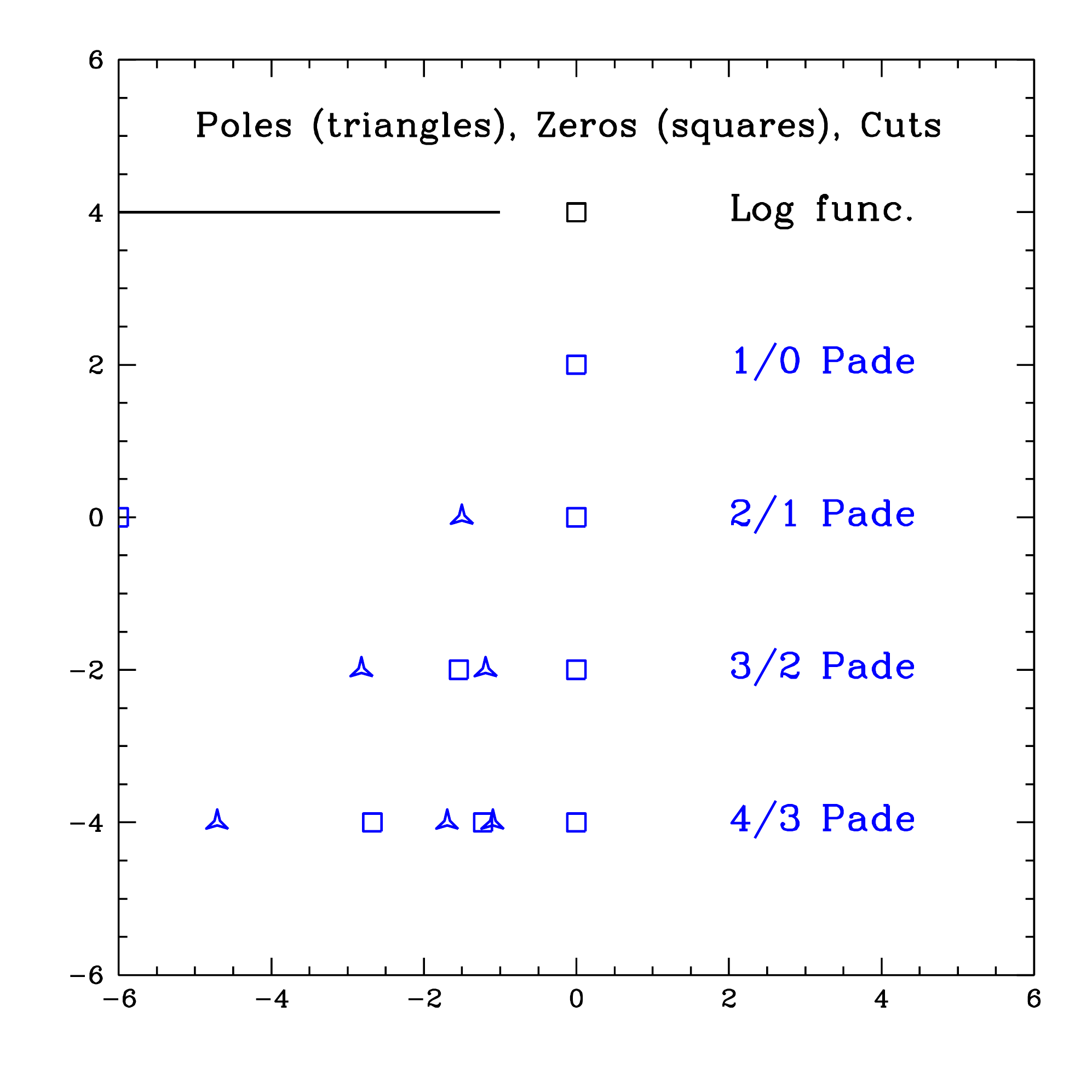}
    \caption{\label{figLog}
      Fitting the logarithm function $\ln(1+x)$ with Pad\'e.  Left:
      comparing the log function to the (1,0), (2,1), (3,2), and (4,3)
      Pad\'e approximants.  The higher approximants do an excellent
      job above $x=-1$ and have a series of poles and zeros where the
      logarithm has its cut.  Right:  the location of zeros and poles
      of the Pad\'e approximants, compared to the zero and cut of
      $\ln(1+x)$.}
\end{figure}

Figure \ref{figLog} shows that such Pad\'e approximants (the figure
considers $(M+1,M)$ approximants) work very well near $x=0$ (over a
much wider range than the Taylor series, not shown), but ``go crazy''
where $\ln(1+x)$ has its cut.  Since the Pad\'e function cannot have a
cut, it instead has an alternating series of zeros and poles.  As one
uses more terms to get a better approximant, the poles and cuts get
closer together and also cover more of the cut.  If we could take the
limit of a large order $M$ of the Pad\'e approximant, we would see the
zeros and poles always alternating but getting closer and closer
together and filling more and more of the negative axis.

\begin{figure}[ht]
  \includegraphics[width=0.46\textwidth]{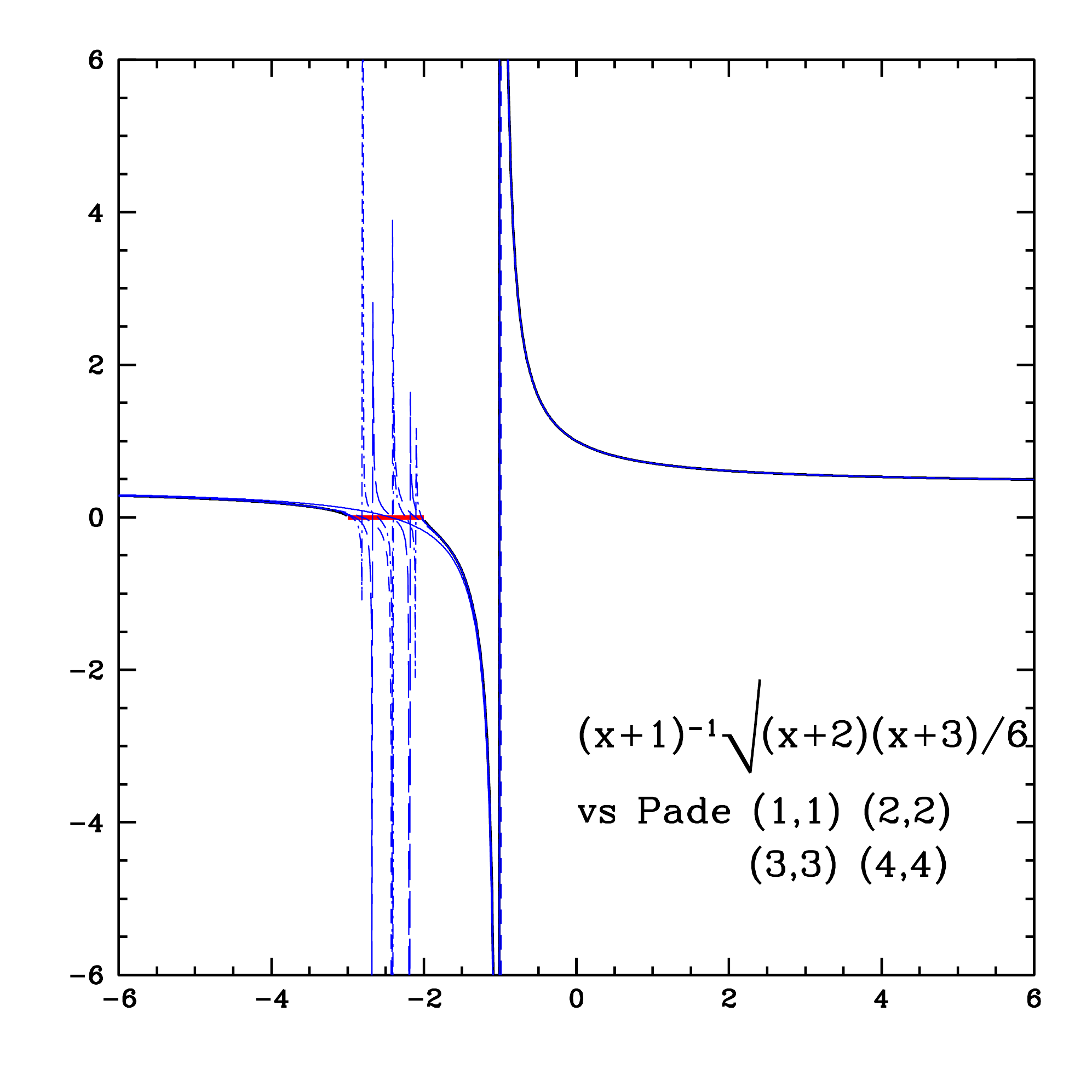}
  \hfill
  \includegraphics[width=0.46\textwidth]{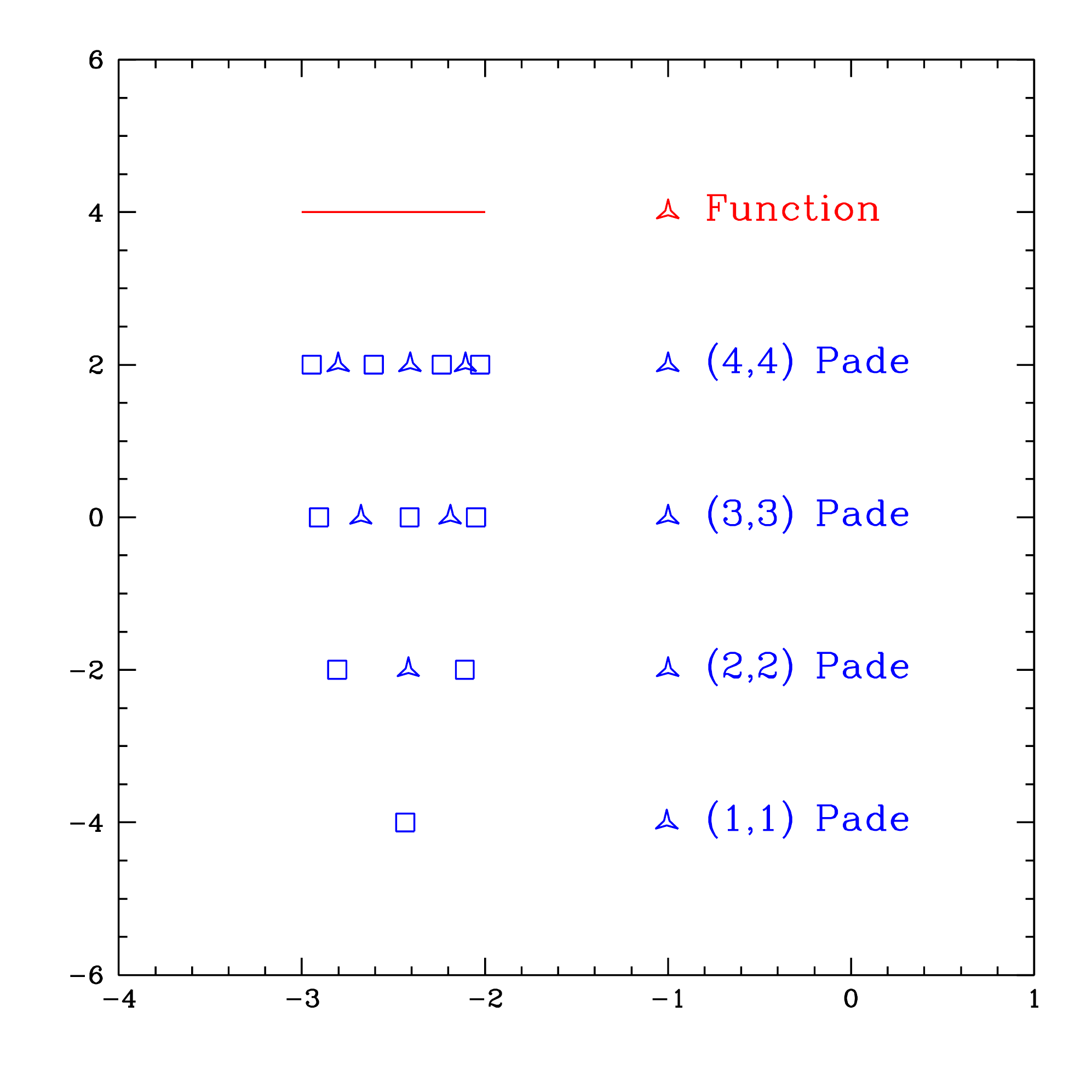}
  \caption{\label{figsqrt}
    Pad\'e fits (left) and poles/zeros (right) for the function
    $\sqrt{(x+2)(x+3)/6}/(x+1)$, which has a pole and a cut.}
\end{figure}

We can also distinguish a pole from a cut in a function which has
both; Figure \ref{figsqrt} shows a similar study of the function
$f=\sqrt{(x+2)(x+3)/6} / (x+1)$, which has a pole at $x=-1$ and a cut
from $x=-2$ to $x=-3$.  The Pad\'e approximants fit the function well
everywhere but in the cut, and the cut is easily identified as a
region where alternating zeros and poles get denser and denser.
The leading pole is fit very accurately, for essentially the same
reason that the variational method is so good at establishing the
ground state energy in quantum mechanics.  We also studied a function
with a series of poles, though we do not provide a figure.  In this
case, the Pad\'e approximant comes close to capturing the first few
poles but misses those which are farther away.  This is again similar
to how a multi-parameter variational solution finds the first few
energy levels of a quantum system but does a poor job with higher
states.

\subsection{Application: shear correlator}

\begin{figure}[ht]
  \includegraphics[width=0.4\textwidth]{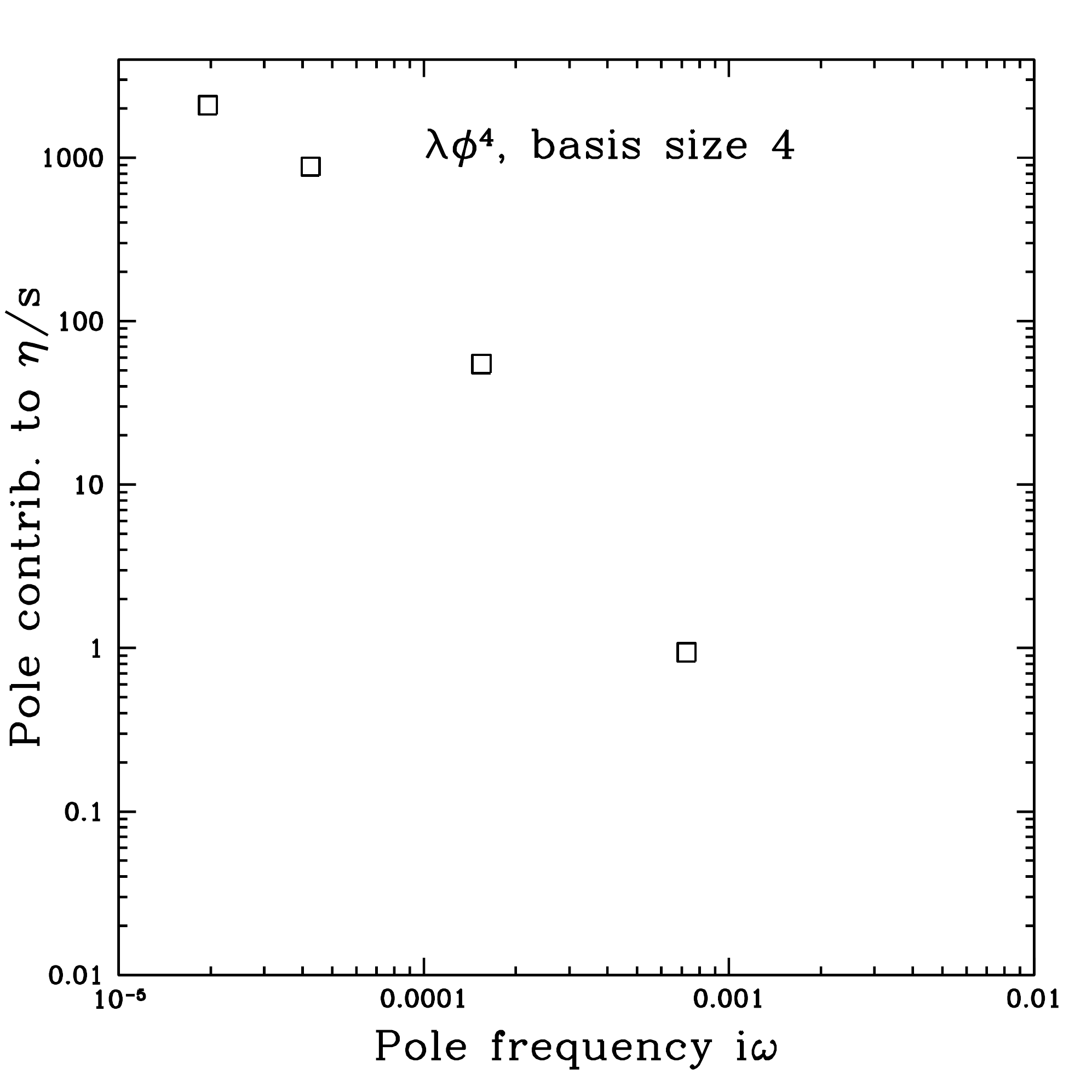}
  \hfill
  \includegraphics[width=0.4\textwidth]{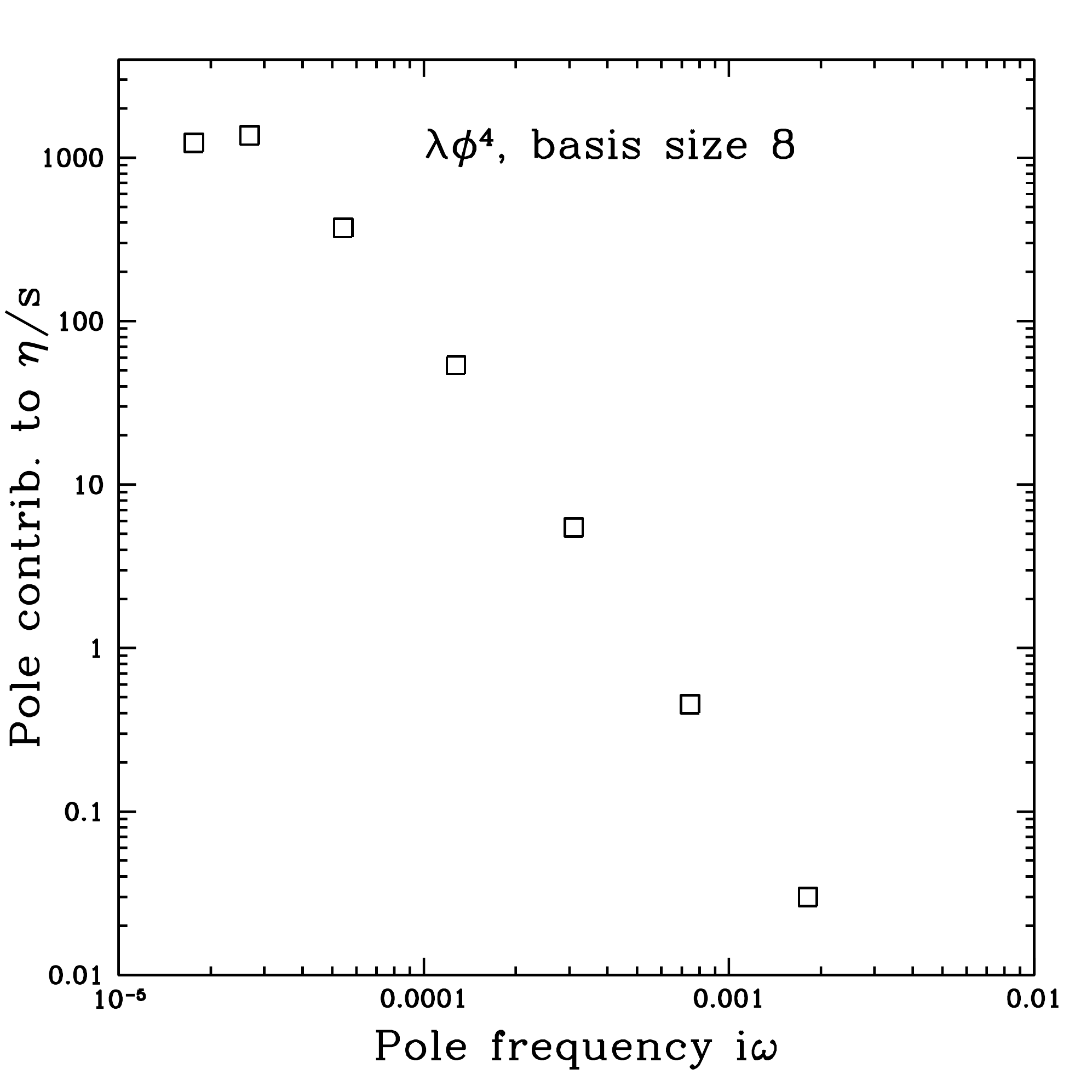}
  \hfill $\phantom{.}$
  \includegraphics[width=0.4\textwidth]{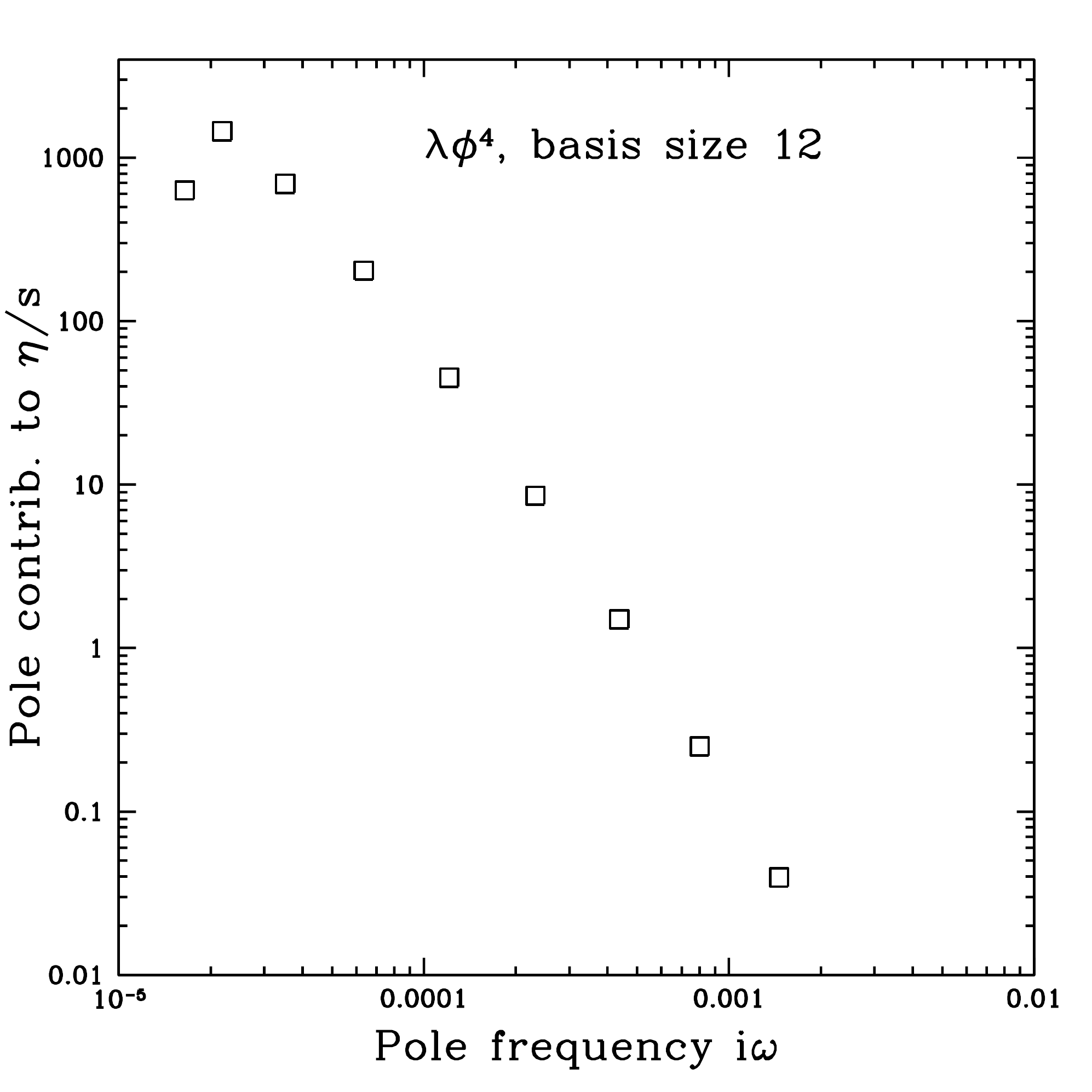}
  \hfill
  \includegraphics[width=0.4\textwidth]{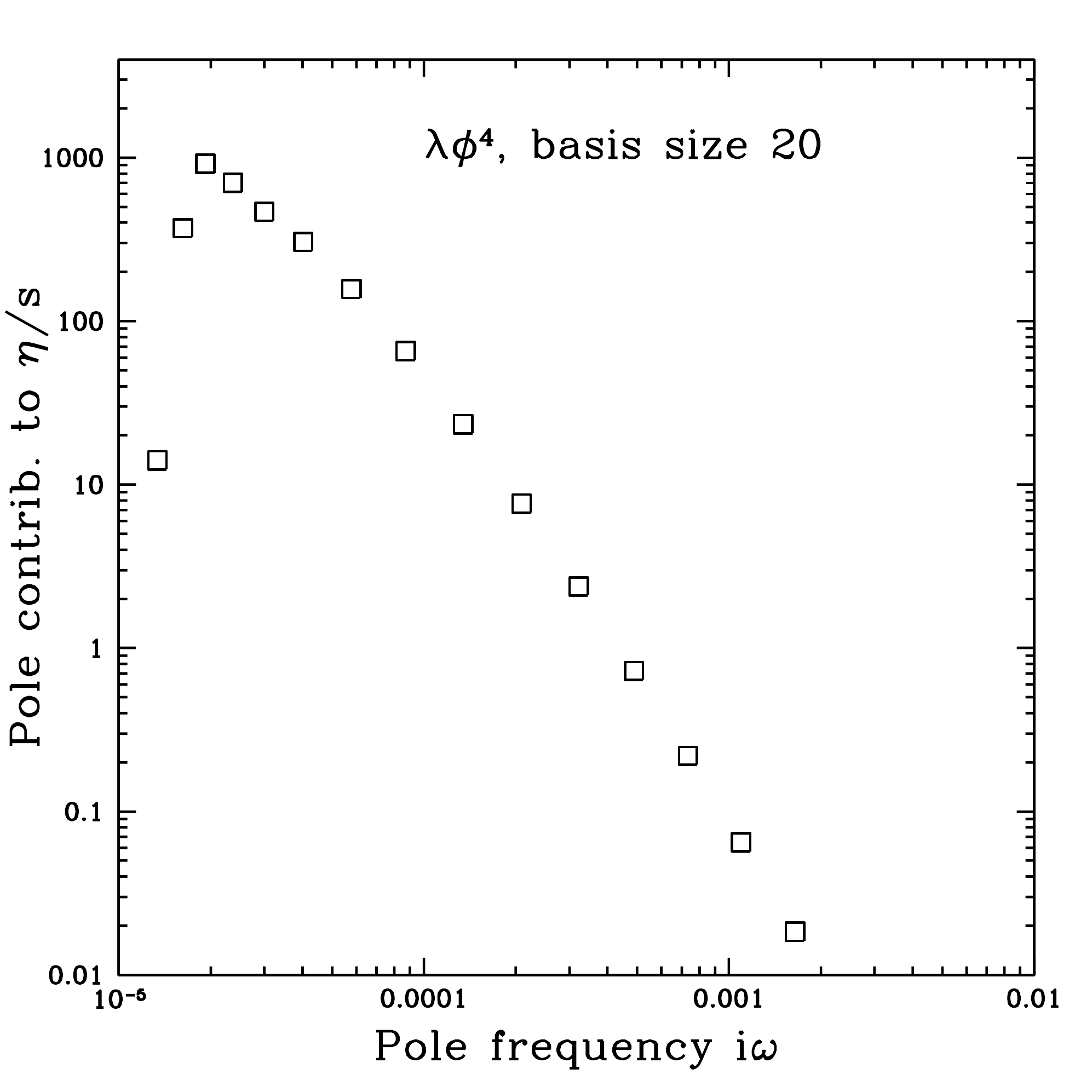}
  \hfill $\phantom{.}$
  \caption{\label{figpoles}
    Location of poles, and contribution of each pole to $\eta$, for
    $\GTT$ in $\lambda \phi^4$ theory as a function of the size of the
    variational \textsl{Ansatz}.  From top left to bottom right, the
    figures represent $(M,N)$ of $(4,4)$, $(10,8)$, $(16,12)$, and
    $(20,16)$.  With increasing basis size, the poles span a wider
    range but also draw closer together.}
\end{figure}

With this in mind, we present our results for the locations of poles
in the stress-stress correlator.  Rather than plot the locations of
poles and of zeros,
which continue to interleave, we plot only the locations of poles, but
we plot them against each eigenvector's contribution to the shear
viscosity as found in \Eq{eta_evec} -- that is, in each plot we use as
the $y$-axis the value
$\frac{T^2}{5\lambda_i} |\langle \chi(p,0)\,|\,\xi_i\rangle|^2$
which the eigenvalue contributes to the shear viscosity or
equivalently the residue of the pole in the viscosity correlator.

We start in Figure
\ref{figpoles} by showing the dependence of the pole location and
shear-contribution on the size of the variational basis.  Each of the
four panels represents the choice $p_0=T$, but we choose $(M,N)$ (the
size of the
basis and the power in the denominator) to be $(4,4)$, $(10,8)$,
$(16,12)$, and $(20,16)$.  As the figure shows, increasing the basis
size leads to poles which cover a
wider frequency range but are also denser, just as we found in
the examples where a function with poles is approximating a cut.  The
functional form of \Eq{G_evec} ensures that zeros interleave between
these poles.  The lowest pole also gets progressively lower with
smaller residue, rather than becoming fixed, indicating that the first
nonanalytic feature is not a pole but a branch point with very small
initial discontinuity.  (If the true behavior is a cut, then a large
basis will have narrowly-spaced poles and the cut discontinuity will
be the ratio of the pole residue to the inter-pole spacing.)
Note that each figure is a log-log plot; the
range of frequencies and of contributions to $\eta$ are very large.
The $x,y$ axes are missing some units; $\omega$ is measured in units
of $\lambda^2 T$ and $\eta/s$ has a missing factor of $1/\lambda^2$.

\begin{figure}[ht]
\centerline{
  \includegraphics[width=0.6\textwidth]{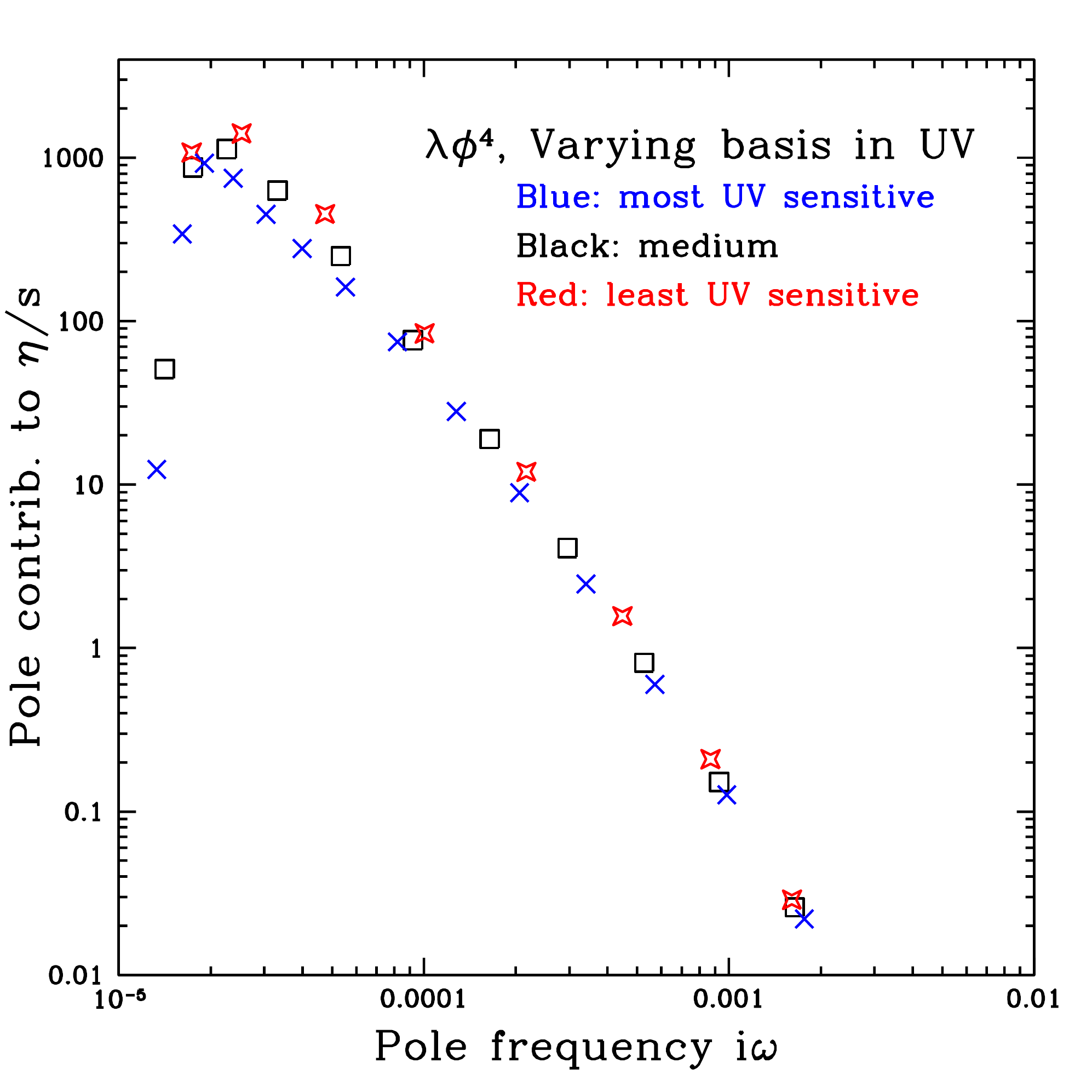}}
\caption{\label{figsens}
  Pole locations for fixed $N=12$ and $M=12,14,16$, with
  $p_0 = 0.5, 1, 2$.  As basis
  functions are added at large $p$ and the scale of sensitivity is
  shifted towards large $p$, pole locations move and the lowest
  observed frequency gets lower.}
\end{figure}

If the spectrum contains poles and not a cut, one would also expect
the locations of the more IR poles which contribute the most to $\eta$
to be stable to changes in the basis details.  This is not the case,
as shown in Figure \ref{figsens}.  The figure shows what happens as we
increase $M$ at fixed $N$, which from \Eq{chiAnsatz} means that we are
adding basis functions onto the UV end of the $p$-spectrum.
Specifically, each pole set in the plot was found with a functional
basis with $N=12$, but $M$ was increased from $12$ to $14$ to $16$
while simultaneously changing $p_0$ from $0.5$ to $1$ to $2$.
Therefore the functional basis became more sensitive to large-$p$
structure, but somewhat less sensitive to small-$p$ (IR) structure.
As we shift from more IR to more UV sensitivity, the
location of poles shifts, the poles get closer together (bringing down
the individual contributions to $\eta$ so that the sum stays the
same), and the smallest $\omega$ value observed gets smaller.
(Note that 4, 3, and 2 poles lie off the high-frequency, small-residue
side of the plot for the most IR, medium, and most UV sensitive basis,
indicating that the most IR sensitive basis has more sensitivity at
the high-frequency end.)
This shows that the lowest-$\omega$ poles correspond to functional
forms which lie mostly at large $p$, while the highest-$\omega$ poles
lie mostly at small $p$.  This behavior makes sense because the
total scattering cross-section in $\lambda \phi^4$ theory scales as
$1/s$ and therefore as $1/p$, indicating that high-energy particles
live longer and low-energy particles change more quickly.

\begin{figure}[ht]
\centerline{
  \includegraphics[width=0.6\textwidth]{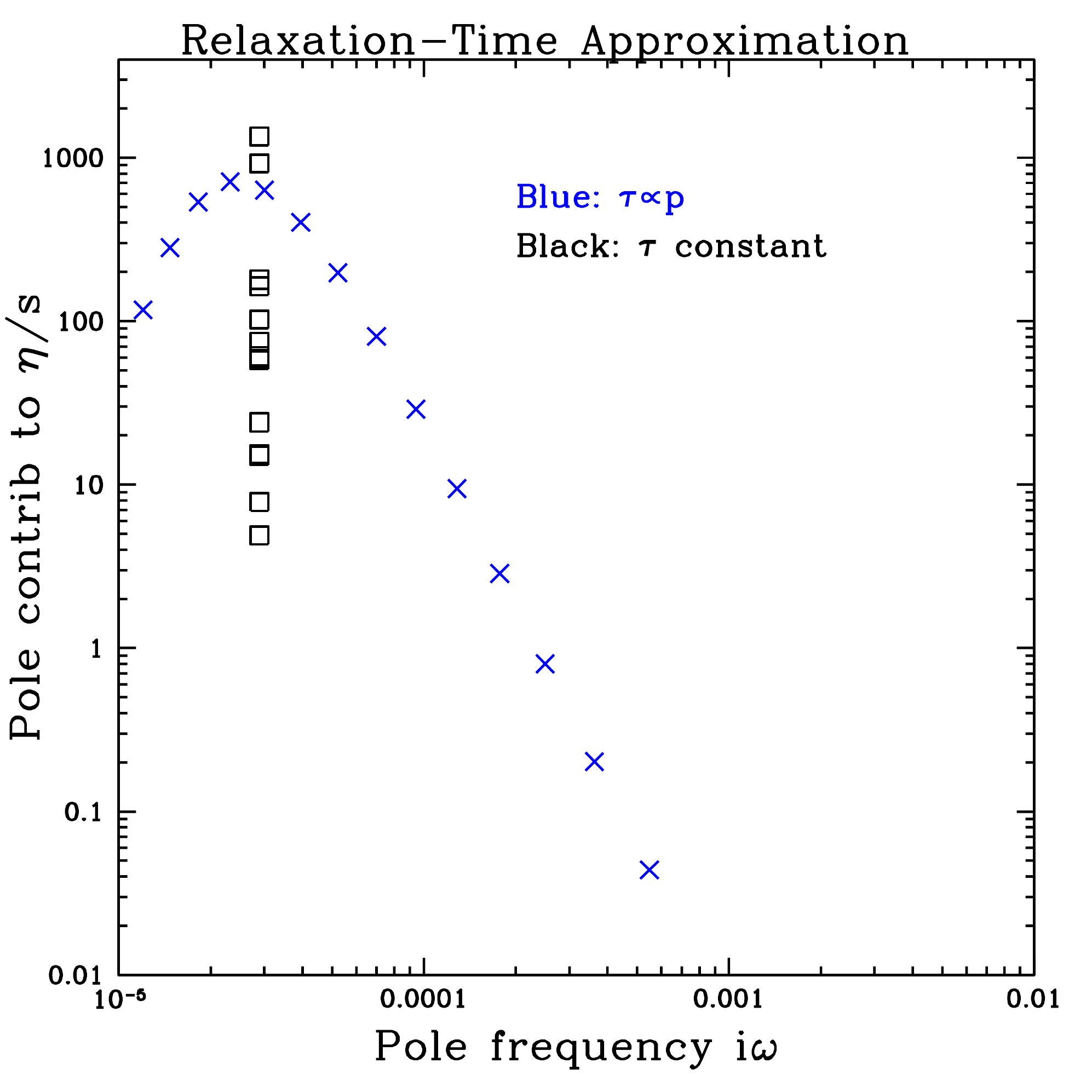}}
\caption{\label{figrelax}
  Pole locations when we apply the variational \textsl{Ansatz} method
  to the collision operator $\C$ in the relaxation-time
  approximation, with a fixed relaxation time or with
  $\tau \propto p$ (each scaled to give the same $\eta$ result as the
  full case).}
\end{figure}

Finally it is useful to compare these results to what we would find in
a simpleminded approximation for $\C$ where we know the analytical
structure.  Consider then the generalized momentum-dependent
relaxation-time approximation
\begin{equation}
  \C[\chi(p)] = \tau^{-1}[p] \; \chi(p) \, ,
  \qquad
  \tau[p] = \tau_0 (T/p)^\alpha \,,
\end{equation}
where $\alpha$ controls the momentum dependence of the relaxation.
The case $\alpha=0$ is the traditional relaxation-time approximation,
in which all excitations approach equilibrium with the same speed.
The choice $\alpha=-1$ or $\tau \propto p$ means that UV excitations
relax more slowly
than IR ones.  This is the closest one can come to the real behavior
of $\lambda \phi^4$ theory within this family of approximations.
Recently Kurkela and Wiedemann have explored the analytic structure of
the $\langle T^{0i} T^{0i} \rangle(k)$ (shear-channel) correlator as a
function of $k$ in this approximation \cite{Kurkela:2017xis}, in a
study complementary to ours.  Romatschke's similar study
\cite{Romatschke:2015gic} considers only $\alpha=0$, so $\tau=\tau_0$
independent of $p$.

Within this simplified approximation, the collision operator
$\C[\chi(p)]$ is diagonal in the $p$-basis, so the eigenvectors are
the functions $\chi(p) \propto \delta(p-p_0)$, indexed by $p_0$.
For $\alpha=0$ the eigenvalues are $\omega=1/\tau_0$ for all $p_0$.
This degenerate eigenspectrum means that we can choose one eigenvector
with perfect projection against $|\, \chi(p,0)\rangle$, leading to a
spectrum with a single pole.  For $\alpha=-1$ the eigenvalues are
$\omega=T/(p_0 \tau_0)$, leading to a cut along the whole imaginary
$\omega$ axis.  Since small $\omega$ corresponds to large $p_0$, the
cut has exponentially small discontinuity for small $\omega$; whereas
for large $\omega$ and therefore small $p_0$, the discontinuity decays
as a power of $\omega$.  We might expect a similar behavior in
$\lambda \phi^4$ theory because the total scattering cross-section
scales as $\sigma \sim 1/s \sim 1/p$.

We can use this solvable case to test how our multi-parameter
\textsl{Ansatz} method performs.  The results are shown
in Figure \ref{figrelax}.  In each case
we have set the overall coefficient of $\C$ to reproduce the same
$\eta$ value that we found in the complete treatment, so Figures
\ref{figpoles}, \ref{figsens}, and \ref{figrelax} are directly
comparable.%
\footnote{To make Figure \ref{figrelax} we used $N=16$ and $M=20$ with
  $p_0=T$.}
The figure shows that the single relaxation time indeed finds a single
frequency.  Our eigenvalue solver happened to pick a basis where the
spectral weight was distributed over several modes, but since the
modes are degenerate this is an arbitrary choice and we could change
the basis to find a single pole.  On the other hand,
for the $p$-dependent relaxation-time approximation we find a series
of poles which look quite similar to the behavior found for the true
collision operator.  That is, when we consider two toy collision
operators, one with a pure-point spectrum and one with a continuous
spectrum, the actual behavior of $\lambda \phi^4$ closely resembles
the continuous-spectrum toy model when we view each using the
\textsl{Ansatz} method.

\section{Discussion}

The correlator of the trace-subtracted stress tensor plays a central
role in hydrodynamics, controlling both the shear viscosity and
certain ``second-order'' transport coefficients.  If hydrodynamics is
to be extended from describing slowly-varying systems near local
equilibrium to more accurately reflecting microscopic relaxation
processes (something we can hope for in second-order hydro), then we
need to understand the analytical structure of $\GTT(\omega)$ as
accurately as possible.

Does $\GTT(\omega)$ feature poles, as in strongly-coupled theories
with holographic duals and in the simplest relaxation-time
approximation?  Or does it feature a cut or more complex analytic
structure?  We explored this question in a weakly-coupled theory,
$\lambda \phi^4$ theory.  Within the kinetic-theory approximation
(which should describe most weakly-coupled theories), the correlator
has its nonanalyticities strictly on the negative imaginary-frequency
axis.  And in $\lambda \phi^4$ theory, the nonanalyticity appears to
be a cut.  The evidence for a cut is that when we render the problem
solvable, by restricting to a finite-dimensional (but large) subspace
of departures from equilibrium, we find a dense set of poles which
grow denser as the subspace is expanded.  This behavior is typical for an
approximation which must find poles, used on a function which in truth
has a cut.  It is also what we find in the $p$-dependent
relaxation-time approximation, where we know there is a cut.  And like
the $p$-dependent relaxation-time approximation, we believe that the
cut runs all the way up to $\omega=0$, albeit with exponentially
shrinking discontinuity.

We could attempt to make the same study for weakly coupled QCD.
The collision integral is also known \cite{Arnold:2003zc},
but its more complicated form means that we cannot integrate it with
enough precision to use really large bases of test functions.
Therefore it is not so easy to see the limiting behavior as the basis
is made large.  However we anticipate the same qualitative behavior.
In fact, since processes which exchange a small momentum have a large
cross-section in gauge theories, we expect a much larger ``tail''
towards much higher frequencies within QCD, reflecting $\chi(p)$
functions which oscillate rapidly as a function of $p$.

The most important conclusion of our work is that it is perfectly
possible, indeed should perhaps be expected, that the stress-stress
correlator $\GTT(\omega)$ in real-world QCD has an analytic structure
which is more complicated than a few well-isolated poles.  Even if we
knew this structure -- for instance, if we assume that it is the same
as in $\lambda \phi^4$ theory -- this precludes a second-order
hydrodynamical treatment such as \Eq{IS2} from serving as a really
accurate \textsl{microphysical} description of the relaxation process.

What we have \textsl{not} done is to study the behavior of the
stress-tensor correlator at finite wave number $k$.  To do so with the
full nontrivial collision operator would require an extension of this
work from departures from equilibrium with $\ell=2$ spherical-harmonic
structure, to those with all $\ell$ values.  Such an approach would
answer the question of how realistic collision integrals affect the
conclusions of Romatschke's \cite{Romatschke:2015gic} and
Kurkela and Wiedemann's \cite{Kurkela:2017xis} work.  While
technically feasible, such a study would be significantly more
complicated, so we leave it for future work.

\section*{Acknowledgments}

We thank the Technische Universit\"at Darmstadt and its Institut f\"ur
Kernphysik, where this work was conducted.  We also thank Rolf Baier
and Andrei Starinets for inspiring us to consider this problem, and to
Paul Romatschke and Aleksi Kurkela for valuable discussions.  The
author acknowledges support by the Deutsche Forschungsgemeinschaft
(DFG) through the grant CRC-TR 211 ``Strong-interaction matter under
extreme conditions.''

\bibliographystyle{unsrt}
\bibliography{refs}

\end{document}